\documentclass[conference,a4paper,10pt]{IEEEtran}

\IEEEoverridecommandlockouts

\usepackage[final]{graphicx}
\usepackage{times}
\usepackage{amsfonts}
\usepackage{cite}
\usepackage{citesort}
\usepackage{epsfig}
\usepackage{amssymb}
\usepackage{amsmath}
\usepackage{psfrag}
\usepackage{xcolor}
\usepackage{cite}
\usepackage{amsbsy}
\usepackage{verbatim}
\usepackage{ctable}

\usepackage{mathtools}

\newcommand{\otoprule}{\midrule[\heavyrulewidth]}

\newcommand\coolunder[2]{\mathrlap{\smash{\underbrace{\phantom{%
    \begin{matrix} #2 \end{matrix}}}_{\mbox{$#1$}}}}#2}

\newtheorem{lemma}{Lemma}
\newtheorem{corollary}{Corollary}
\newtheorem{definition}{Definition}
\newtheorem{theorem}{Theorem}

\newtheorem{proposition}{Proposition}

\def\thBP{\varepsilon^{\mathrm{BP}}}

\def\dl{d_{\mathrm v}}
\def\dr{d_{\mathrm c}}
\def\pvec{{\boldsymbol p}}
\def\xvec{{\boldsymbol x}}
\def\avec{{\boldsymbol a}}

\def\avecc{{\boldsymbol a}_\circ}
\def\bvecc{{\boldsymbol b}_\circ}

\def\Fvec{{\boldsymbol F}}
\def\Gvec{{\boldsymbol G}}
\def\Xvec{{\boldsymbol X}}
\def\Yvec{{\boldsymbol Y}}
\def\yvec{{\boldsymbol y}}

\def\Dvec{{\boldsymbol D}}
\def\H{{\boldsymbol H}}
\def\Avec{{\boldsymbol A}}

\def\fvec{{\boldsymbol f}}

\def\gvec{{\boldsymbol g}}

\def\zerovec{{\boldsymbol 0}}
\def\pvecO{{\boldsymbol p}_{\circ}}
\def\xvecO{{\boldsymbol x}_{\circ}}
\def\yvecO{{\boldsymbol y}_{\circ}}
\def\fvecO{{\boldsymbol f}_{\circ}}
\def\gvecO{{\boldsymbol g}_{\circ}}

\newcommand{\set}[2]{\mathcal{S}^{#1}_{#2}}

\newcommand{\setf}[1]{\mathcal{S}^{f}_{#1}}

\newcommand{\derivF}[2]{\frac{\partial #1}{\partial #2}}
\def\Egap{{\Delta E}}
\newcommand{\T}{^{\mathsf{T}}}

\def\dw{{\mathtt w}}

\def\tr{{\mathrm{Tr}}}

\begin{document}
\include{macros}

\title{Proving Threshold Saturation for Nonbinary SC-LDPC Codes on the Binary Erasure Channel}

\author{Alexandre Graell i Amat$^{\dag}$, Iryna Andriyanova$^{\ddag}$, Amina Piemontese$^{*}$ \\
\normalsize $\dag$ Department of Signals and Systems, Chalmers University of Technology, Gothenburg, Sweden\\
$\ddag$ETIS Laboratory, ENSEA/University of Cergy-Pontoise/CNRS, Cergy-Pontoise, France\\
$*$Universit\`a di Parma, Dipartimento di Ingegneria dell'Informazione, Viale G. P. Usberti 181/A, Parma, Italy
\thanks{Research supported by the Swedish Research Council under grant \#2011-5961, the Swedish Foundation for Strategic Research (SSF) under the Gustaf Dal\'en project IMF11-0077, and the Adlerbert Research Foundation.}}
\maketitle

\begin{abstract}
We analyze nonbinary spatially-coupled low-density parity-check (SC-LDPC) codes built on the general linear group for transmission over the binary erasure channel. We prove threshold saturation of the belief propagation decoding to the potential threshold, by generalizing the proof technique based on potential functions recently introduced by Yedla \textit{et al.}. The existence of the potential function is also discussed for a vector sparse system in the general case, and some existence conditions are developed. We finally give density evolution and simulation results for several nonbinary SC-LDPC code ensembles.
\end{abstract}

\section{Introduction}

Spatial coupling of low-density parity-check (LDPC) codes has revealed as a powerful technique to construct codes that universally achieve capacity for many channels under belief propagation (BP) decoding. The main principle behind this outstanding behavior is the convergence of the BP threshold to the maximum a posteriori (MAP) threshold of the underlying block code ensemble, a phenomenon known as threshold saturation \cite{KuRiUr11}. The concept of spatial coupling is not exclusive of LDPC codes, and also applies to other scenarios, such as relaying, compressed sensing, and statistical physics. In the realm of coding, spatial coupling has also been recently applied to turbo codes \cite{MolLenGra14}. 

Nonbinary LDPC codes defined over GF$(2^m)$ have received an increasing interest in the recent years, since for short-to-moderate block lengths they have been shown to outperform their binary counterparts. Nonbinary spatially-coupled LDPC (SC-LDPC) codes have been considered more recently in, e.g., \cite{PieGraCol13,Hua14}. In \cite{PieGraCol13} a method to compute an upper bound on the MAP threshold for nonbinary LDPC codes on the binary erasure channel (BEC) was proposed, and it was shown that the MAP threshold of regular ensembles improves with $m$ and approaches the Shannon limit. It was also empirically shown in \cite{PieGraCol13} that threshold saturation occurs for nonbinary SC-LDPC codes. 

In this paper, we prove\footnote{The proofs of the main results in this paper can be found in the extended version of the paper \cite{AndGra13}.} threshold saturation of the BP threshold of nonbinary SC-LDPC codes on the BEC to the so called potential threshold, which is conjectured to coincide with the MAP threshold. Our proof is based on the proof technique proposed in \cite{pfister-scalar-arxiv,pfister-itw} to prove threshold saturation for (binary) SC-LDPC codes. This technique is based on the observation that the density evolution (DE) equations of LDPC codes form an admissible system for which it is possible to properly define a potential function, and that a fixed point of the DE corresponds to a stationary point of the corresponding potential function. Our proof is a non-trivial generalization of the proof in \cite{pfister-itw} to accomodate nonbinary SC-LDPC codes. In particular, we discuss the necessary conditions for the existence of the potential function for a vector sparse system in the general case, and show that the potential function in the form of \cite{pfister-itw} does not exist for nonbinary codes. We also give DE results and simulation results for several nonbinary SC-LDPC code ensembles.

\subsection{Notation and Some Definitions}
\label{sec:notation}
We use upper case letters $F$ to denote scalar functions, bold lowercase letters $\xvec$ to denote vectors, and bold uppercase letters $\Xvec$ for matrices. We assume all vectors to be row vectors, and we denote by ${\mathrm{vec}}( \Xvec)$ the row vector obtained by transposing the vector of stacked columns of matrix $\Xvec$.

Let $\xvec=(x_1, \ldots, x_m)$ be a non-negative vector of length $m$.  
The Jacobian of a scalar function $F(\xvec)$ is defined as $F'=\derivF{F(\xvec)}{\xvec}=\left(\derivF{F}{x_1},\ldots,\derivF{F}{x_m}\right)$. Also, we define the Jacobian of a vector function $\fvec$ as $\Fvec_{\mathrm d}(\xvec)=\fvec'(\xvec)=\left(\derivF{\fvec(\xvec; \varepsilon)_k}{x_n}\right)$, where $k=1,\ldots,m$ and $n=1,\ldots,m$, and the Hessian of a vector function $\fvec$ as $\Fvec_{\mathrm{dd}}(\xvec) = \fvec''(\xvec)=\left(\derivF{{\mathrm{vec}}  ( \Fvec_{\mathrm d}(\xvec))_k}{x_n}\right)$.

\label{sec:PotentialRegular}

\section{Density Evolution for $(\dl, \dr, m)$ and $(\dl, \dr, m, L, \dw)$ LDPC Code Ensembles over GF($2^m$)}
\label{sec:DE}
We consider transmission over a BEC with erasure probability $\varepsilon$, denoted by BEC($\varepsilon$), using nonbinary LDPC codes defined over the general linear group. The code symbols are elements of the binary vector space GF$(2^m)$, of dimension $m$, and we transmit on the BEC the $m$-tuples representing their binary image. We denote a regular nonbinary LDPC code ensemble over GF$(2^m)$ as  $(\dl, \dr, m)$, where $\dl$ and $\dr$ denote the variable node degree and the check node degree, respectively. In this paper, we will also consider the regular $(\dl, \dr,m, L, \dw)$ SC-LDPC code ensembles described in \cite{KuRiUr11}, where $L$ denotes the spatial dimension, and $\dw>0$ is the \textit{smoothing} parameter. This ensemble is obtained by placing $L$ sets of variable nodes of degree $\dl$ at positions $t\in\{1,\ldots,L\}$. A variable node at position $t$ has $\dl$ connections to check nodes at positions from the range $[t,t+\dw-1]$. For each connection, the position of the check node is uniformly 
and independently chosen from that range. 
A (terminated) $(\dl, \dr,m, L, \dw)$ SC-LDPC code ensemble is  defined by the parity-check matrix
\begin{equation*}
\label{Eq:Hmatrix}
\small
\H=\left[\begin{array}{c c c  }
 \H_0(1) & &   \\
 \vdots & \ddots &   \\
 \H_{\dw-1}(1) & &   \\
 & &     \H_0(L) \\
 &\ddots &  \vdots  \\
 & &   \H_{\dw-1}(L)
\end{array}\right].
\end{equation*} 
Each submatrix $\H_i(t)$ is a sparse $(M \dl/\dr)\times M$ nonbinary matrix, where $M$ is the number of variable nodes in each position and $M \dl/\dr$ the number of check nodes in each position. It is important to note that the check node
degrees corresponding to the first and last couple
of positions is lower than $\dr$, i.e., the graph shows some irregularities. These irregularities lead to a locally better decoding (at the expense of a rate loss, which vanishes with $L$) and are the responsible for the outstanding performance of SC-LDPC codes.

In general, the messages exchanged in the BP decoding of nonbinary LDPC codes are real vectors of length $2^m$, the $i$th element of which representing the a posteriori probability that the symbol is $i$. For the DE on the BEC, however, it is sufficient to keep track of the dimension of the messages exchanged. Therefore, the DE simplifies to the exchange of messages of length $m+1$, where the $i$th entry of the message is the probability that the message has dimension $i$. 

\subsection{$(\dl, \dr, m)$ Regular LDPC Code Ensemble over GF($2^m$)}
Consider a $(\dl, \dr, m)$ ensemble over GF$(2^m)$, used for transmission over the BEC($\varepsilon$). Let $\xvecO^{\ell}= (x_{\circ 0}^{\ell}, \ldots, x_{\circ m}^{\ell})$ and $\yvecO^{\ell}= (y_{\circ 0}^{\ell}, \ldots, y_{\circ m}^{\ell})$ be probability vectors of length $m+1$, where $x_{\circ i}^{\ell}$ (resp. $y_{\circ i}^{\ell}$) is the probability that a message from (resp. to) variable nodes at iteration ${\ell}$ has dimension $i$, $0 \le i \le m$. The DE updates for the variable nodes and the check nodes at iteration $\ell$ are described by
\begin{align}
\xvecO^{\ell} = \fvecO(\yvecO^{\ell}; \varepsilon),~~~~
\yvecO^{\ell} = \gvecO (\xvecO^{\ell-1}).\nonumber
\end{align}
where $\fvecO=(f_{\circ 0},\ldots,f_{\circ m})$ and $\gvecO=(g_{\circ 0},\ldots,g_{\circ m})$ are 
functions
from $[0,1]^{m+1}$ to $[0,1]^{m+1}$, defined as
\begin{align}
\label{eq:bij1}
\fvecO(\yvecO; \varepsilon)  &= \pvecO(\varepsilon) \boxdot \left( \boxdot^{\dl-1} \yvecO \right),
\\
\label{eq:bij2}
\gvecO(\xvecO) & = \boxtimes^{\dr-1} \xvecO .
\end{align}
For two probability vectors $\avecc$ and $\bvecc$ of length $m+1$, the operations $\boxdot$ and $\boxtimes$ in (\ref{eq:bij1})-(\ref{eq:bij2}) are defined as
\begin{align} 
[\avecc \boxdot \bvecc]_k&=\sum_{i=k}^m \sum_{j=k}^{k+m-1} V^m_{i,j,k} a_{\circ i} b_{\circ j}, \quad k=0,\ldots,m,\label{eq:boxdot}\\
[\avecc \boxtimes \bvecc]_k&= \sum_{i=0}^k \sum_{j=k-i}^{k} C^m_{i,j,k} a_{\circ i} b_{\circ j}, \quad k=0,\ldots,m,\label{eq:boxtimes}
\end{align}
where $V^m_{i,j,k}$ is the probability of choosing a subspace of dimension $j$ whose intersection with a subspace of dimension $i$ has dimension $k$, and $C^m_{i,j,k}$ is the probability of choosing a subspace of dimension $j$ whose sum with a subspace of dimension $i$ has dimension $k$ (see \cite{AndGra13} for details). Moreover, we define $\boxdot^{\dl-1} \avecc = \avecc \boxdot \avecc \boxdot \cdots \boxdot \avecc$ with $\dl-1$ terms $\avecc$ (i.e., $\boxdot^1 \avecc = \avecc$), and
$\boxtimes^{\dl-1} \avec = \avecc \boxtimes \avecc \boxtimes \cdots \boxtimes \avecc$ with $\dl-1$ terms $\avecc$ (i.e., $\boxtimes^1 \avecc = \avecc$).

In (\ref{eq:bij1}) $\pvecO$ is a row vector of length $m+1$, the $i$th element of which being the probability that the channel message has dimension $i$,
\begin{equation}
\label{eq:pvecO}
{\pvecO}_{i}(\varepsilon) = \binom{m}{i} \varepsilon^i (1-\varepsilon)^{m-i} , \; i=0,\cdots, m\, .
\end{equation}
Also, 
$\xvecO^{0}=\pvecO$. 

The fixed-point DE equation for $\xvecO = \xvecO^{\infty}$ is
\begin{equation}
\label{eq:DElrm0}
\xvecO = \fvecO(\gvecO(\xvecO);\varepsilon).
\end{equation}
Note that decoding is successful when the DE equation converges to $\xvecO^{\infty}=(1,0,\ldots,0)$. 

The proof technique in \cite{pfister-itw} requires monotone vector functions for the variable node and the check node updates. It can be shown that $\fvecO$ and $\gvecO$ are not monotone, hence cannot be used directly. In the following, we rewrite the DE equation in (\ref{eq:DElrm0}) in a more suitable form to prove threshold saturation.

\begin{definition}
Given a probability vector $\xvecO$, define the CCDF vector $\xvec=(x_1,\ldots,x_m)$, where $x_i=\sum_{k=i}^m x_{\circ k}$. We also define $x_{m+1}=0$. Then, it follows that $x_{\circ i}=x_i-x_{i+1}$. Note also that $x_{0}=1$. For simplicity of further notation, let $\xvec^{-1}=(1,x_1,\ldots,x_{m-1})$ denote a right shift of $\xvec$ with a prepended 1.
\end{definition}

Considering the CCDF vectors $\xvec$, $\yvec$ and $\pvec$, we can define new vector functions $\fvec(\yvec;\varepsilon)$ and $\gvec(\xvec)$, with
\begin{align*}
f_i &=\sum_{k=i}^m f_{\circ k}(\yvecO;\varepsilon)=\sum_{k=i}^m\left[(\pvec^{-1}-\pvec)\boxdot(\boxdot^{\dl-1}(\yvec^{-1}-\yvec))\right]_k\\
g_i &=\sum_{k=i}^m g_{\circ k}(\xvecO)=\sum_{k=i}^m \left[\boxtimes^{\dr-1}(\xvec^{-1}-\xvec)\right]_k.
\end{align*}

Then, the DE equation (\ref{eq:DElrm0}) can be written in an equivalent form as
\begin{equation}
\label{eq:DElrm}
\xvec = \fvec(\gvec(\xvec);\varepsilon).
\end{equation}
\begin{theorem}
\label{The:Monotonicity}
The functions $\fvec(\xvec; \varepsilon)$ and $\gvec(\xvec)$ are 
increasing in $\xvec$.
\end{theorem}
\begin{corollary}
The density evolution for regular nonbinary LDPC codes given by (\ref{eq:DElrm}) converges to a fixed point. 
\end{corollary}

Successful decoding corresponds to convergence of the DE equation (\ref{eq:DElrm}) to the fixed point $\xvec^{\infty}={\mathbf 0}=(0,0,\ldots,0)$. 

For later use, we denote by $\cal X$ the set of all possible values of $\xvec$. Likewise, we denote by $\cal Y$/$\cal E$ the set of all possible values of $\yvec$/$\varepsilon$. For nonbinary codes and for some $\varepsilon$,
\begin{align}
{\cal E}&:~0 \le \varepsilon \le 1,\nonumber\\
{\cal X}&:~0 \le x_i \le \pvec_{\varepsilon,i},~~~ 
{\cal Y}:~0 \le y_i \le 1,~~\ 1 \le i\le m.\nonumber
\end{align}

Vector functions $\fvec$ and $\gvec$ have several properties which will be useful for the proof of threshold saturation in Section~\ref{sec:PotentialRegular}.
{\lemma\label{lem:Lemma1}
Consider 
$\fvec(\yvec; \varepsilon)$ 
and $\gvec(\xvec)$ 
defined above. For $\xvec\in{\cal X}$ and $\yvec\in{\cal Y}$,
\begin{enumerate}
\item $\fvec(\yvec; \varepsilon)$ and $\gvec(\xvec)$ are nonnegative vectors;
\item $\fvec(\yvec; \varepsilon)$ is differentiable in $\yvec$ and $\gvec(\xvec)$ is twice differentiable in $\xvec$;
\item $\fvec(\zerovec; \varepsilon) = \fvec (\yvec; 0)=\gvec(\zerovec)=\zerovec$;
\item 
$\Gvec_{\mathrm d}(\xvec)> 0$, 
and it is invertible for $\xvec\in{\cal X}\backslash\{\zerovec\}$;
\item $\fvec(\yvec; \varepsilon)$ is strictly increasing with $\varepsilon$.
\end{enumerate}
}

\subsection{$(\dl, \dr,m, L, \dw)$ SC-LDPC Code Ensemble over GF($2^m$)}
Assume a $(\dl, \dr,m, L, \dw)$ ensemble over GF$(2^m)$ and transmission over the BEC($\varepsilon$).  
In the form of (\ref{eq:DElrm}), the fixed-point DE equations for the $(\dl, \dr,m, L, \dw)$ ensemble can be written as 
\begin{align}
\xvec_{i} = \frac{1}{\dw} \sum_{k=0}^{\dw-1} \fvec(\yvec_{i-k};\varepsilon_{i-k}),~~~~
\yvec_{i}= \frac{1}{\dw} \sum_{j=0}^{\dw-1}\gvec(\xvec_{i+j-k}),\nonumber
\end{align} 
where $1 \le i <L+\dw$, and 
$$\varepsilon_i = \begin{cases} \varepsilon,& 1 \le i \le L\\ 0,&1 \le i-L <\dw \end{cases}.
$$

Collect all CCDF vectors $\xvec_i$ and $\yvec_i$ into the $(L +\dw-1)\times m$ matrices 
$\Xvec = (\xvec_1\T, \ldots, \xvec_{L+\dw-1}\T)\T$ and $\Yvec = (\yvec_1\T, \ldots, \yvec_{L+\dw-1}\T)\T$, respectively.
Also, let $\Avec$ be the $L \times (L+\dw-1)$ matrix
\vspace{-1.1cm}
 \[ \vphantom{
    \begin{matrix}
    \overbrace{XYZ}^{\mbox{$R$}}\\ \\ \\ \\ \\ \\ 
    \underbrace{pqr}_{\mbox{$S$}}
    \end{matrix}}%
\begin{matrix}
\vphantom{a}\\ 
\end{matrix}%
\Avec=\frac{1}{\dw}\begin{pmatrix}
1 & 1 & \cdots & 1 & 0 & 0 & \cdots & 0\\
 0&1&\cdots &    1& 1& 0& \cdots &0\\
 \vdots&\vdots&\ddots &    \vdots& \vdots& \vdots& \ddots &\vdots\\
\coolunder{\dw}{0 & 0 & 0 & 0 &} \coolunder{L-1}{0 & 1 & \cdots & 1}\\
\end{pmatrix}%
\begin{matrix}
\end{matrix}\]
\vspace{-0.7cm}
 
The fixed-point DE equation for the $(\dl, \dr,m, L, \dw)$ ensemble can be written in matrix form, similarly as in \cite{pfister-itw},
$$\Xvec = \Avec\T \Fvec (\Avec \Gvec(\Xvec); \varepsilon),$$
where $\Fvec(\Yvec;\varepsilon)$ is an $L\times m$ matrix, $\Fvec(\Yvec;\varepsilon) = (\fvec(\yvec_1;\varepsilon)\T, \ldots, \fvec(\yvec_L;\varepsilon)\T )\T$, $\Gvec(\Xvec)$ is an {$(L+\dw-1)\times m$ matrix, and $\Gvec(\Xvec) = (\gvec(\xvec_1)\T, \ldots, \gvec(\xvec_L+\dw-1)\T )\T$. }


\section{Potential Function and a Proof of Threshold Saturation}
\label{sec:PotentialRegular}
The DE equation (\ref{eq:DElrm}) for the $(\dl,\dr,m)$ regular ensemble describes a vector admissible system for which we can properly define a potential function, similarly to \cite{pfister-itw}.
\begin{definition}
\label{def:PotentialFunction}
The potential function $U(\xvec; \varepsilon)$ of the system defined by functions $\fvec$ and $\gvec$ above,  is given by
\begin{align}
\label{eq:PotentialFunction}
U(\xvec; \varepsilon) 
= \gvec(\xvec)\Dvec \xvec\T -G(\xvec)-F(\gvec(\xvec);\varepsilon),
\end{align}
where $F: {\cal X} \times {\cal E} \mapsto {\mathbb R}$ and $G: {\cal Y} \times {\cal E} \mapsto {\mathbb R}$ are scalar functions that satisfy
$F(\zerovec)=0$, $G(\zerovec)=0$, $F'(\yvec; \varepsilon) = \fvec(\yvec;\varepsilon) \Dvec$, and
$G'(\xvec) = \gvec(\xvec) \Dvec$,
for a symmetric $m\times m$ matrix $\Dvec$ with positive elements $d_{ij}$ and a non-zero determinant.
\end{definition}

The definition of $U(\xvec; \varepsilon)$ above is slightly more general than the one in \cite{pfister-itw}, since $\Dvec$ is assumed to be symmetric with non-zero determinant,  instead of being diagonal as in \cite{pfister-itw}. The properties of $\Dvec$, and the calculation  of $F(\yvec; \varepsilon)$ and $G(\xvec)$ are addressed in Section~\ref{Sec:DefD_F_G}.

{
{\definition 
For $\xvec \in {\cal X}$ and $\varepsilon \in {\cal E}$,
$\xvec$ is a fixed point of the DE if $\xvec = \fvec(\gvec(\xvec);\varepsilon)$; 
$\xvec$ is a stationary point of the potential function if $U'(\xvec;\varepsilon)=\zerovec$.
}

Let the fixed point set be defined as $${\cal F} = \{(\xvec;\varepsilon) | \xvec=\fvec(\gvec(\xvec);\varepsilon)\}.$$

{\lemma
\label{Lem:Potential}
For the vector system defined by $\fvec$ and $\gvec$, the following assertions hold.
\begin{enumerate}
\item $\xvec \in {\cal X}$ is a fixed point if and only if it is a stationary point of the potential $U(\xvec;\varepsilon)$;
\item $U(\xvec;\varepsilon)$ is strictly decreasing in $\varepsilon$, for $\xvec \in {\cal X} \backslash \zerovec$ and $\varepsilon \in \cal{E}$;
\item $U'(\xvec; \varepsilon)$ is strictly decreasing in $\varepsilon$.
\item For some $\varepsilon_1 > 0$ and $\varepsilon_2 > 0$ such that $\varepsilon_1 \neq \varepsilon_2$, if $(\xvec_1, \varepsilon_1) \in {\cal F}$ and $(\xvec_2, \varepsilon_2) \in {\cal F}$, then $\xvec_1\not=\xvec_2$.
\end{enumerate}
}

Thanks to the decreasing property of $U'(\xvec; \varepsilon)$, we can now define the BP and the potential thresholds, denoted respectively by $\thBP$ and $\varepsilon^*$.
{\definition
The BP threshold is
$$\thBP = \sup_\varepsilon \left( \varepsilon \in {\cal E} | U'(\xvec;\varepsilon)>0, \ \forall \xvec \in {\cal X} \right).$$
}
{\definition
The potential threshold is 
$$\varepsilon^* = \sup_\varepsilon \left( \varepsilon \in (\thBP,1] \   | \  \Egap(\varepsilon) \ge 0 \right),$$
where $\Egap(\varepsilon) = \inf_{\xvec \in {\cal X}\backslash {\cal U}_{\zerovec} (\xvec)} U(\xvec; \varepsilon)$, and ${\cal U}_{\zerovec}(\varepsilon) = \{ \xvec \in {\cal X} | \xvec^{\infty} = \zerovec \}.$
}

In other words, $\thBP$ is the lowest value of $\varepsilon$ for which $U(\xvec; \varepsilon)$ does not have a critical point, whereas $\varepsilon^*$ is the lowest value of $\varepsilon$ for which $U(\xvec; \varepsilon)=0$ for all $\xvec$ such that $\xvec^{\infty} \not = \zerovec$.

It has been shown for several systems \cite{pfister-scalar-arxiv}, that the MAP threshold and the potential threshold are identical. We conjecture that the potential threshold of nonbinary LDPC codes is also identical to the MAP threshold.



{
{\definition 
\label{def:PotentialFunctionVec}
The potential function $U(\Xvec;\varepsilon)$ for the spatially-coupled case is defined similarly as in \cite{pfister-itw}
\begin{equation}
\label{eq:PotFunctionVect}
U(\Xvec;\varepsilon) = \tr(\Gvec(\Xvec) \Dvec \Xvec\T ) - G(\Xvec) - F(\Avec \Gvec(\Xvec);\varepsilon),
\end{equation}
where  
{
$G'(\Xvec) = \sum_{i=1}^L G'(\xvec_i) = \sum_{i=1}^L \gvec(\xvec_i) \Dvec$,}
and
{
$F'(\Xvec) = \sum_{i=1}^L F'(\xvec_i) = \sum_{i=1}^L \fvec(\xvec_i) \Dvec$.}
}

The main result of our paper is stated below. It proves successful decoding for $\varepsilon<\varepsilon^*$, i.e., the BP decoder saturates to the potential threshold for large enough values of $\dw$.
{\theorem Consider the spatially-coupled $(\dl, \dr,m,L,\dw)$ LDPC code ensemble, and let $K$ be the upper bound on the norm $||U''(\Xvec;\varepsilon)||_{\infty}$ for its corresponding potential function $U(\xvec; \varepsilon)$.
Then, for $\varepsilon<\varepsilon^*$ and $\dw> \frac{m K}{2 \Egap(\varepsilon)}$, the only fixed point of the system is $\xvec^{\infty}=\zerovec$.
}


{

\section{{Properties of $\Dvec$, and Calculation of $F(\yvec; \varepsilon)$ and $G(\xvec)$}}
\label{Sec:DefD_F_G}

The existence of $F(\yvec; \varepsilon)$ and $G(\xvec)$ is not guaranteed by the definition of $U(\xvec; \varepsilon)$. 
Here, we derive a condition on the existence of $F(\yvec; \varepsilon)$ and $G(\xvec)$ and investigate how it depends on the form of the matrix $\Dvec$. Without loss of generality, we consider the case of the $(\dl, \dr, m)$ ensemble as an example of a coupled vector system.
\begin{theorem}
\label{theorem:existence-F-G}
Consider the $(\dl, \dr, m)$ ensemble
and let $U(\xvec; \varepsilon)$ be given by Definition~\ref{def:PotentialFunction}.
Then, $F(\yvec; \varepsilon)$ and $G(\xvec)$ exist (hence $U(\xvec; \varepsilon)$ exists) if there exist sets of values $\{d_{ j s} \}$, $\{{\varphi}_{(i_1,\ldots,i_m)}\}$ and $\{{\mu}_{(k_1,\ldots k_m)} \}$ that satisfy
\begin{align}
\label{eq:system}
\begin{cases}
i_s {\varphi}_{(i_1,\ldots, i_s, \ldots,i_m)} = \sum_{j=1}^m d_{ j s} {\phi}^{(j)}_{(i_1,\ldots i_s-1, \ldots,i_m)}(\varepsilon)\\
k_t {\mu}_{(k_1,\ldots, k_t, \ldots,k_m)} = \sum_{j=1}^m d_{j t}  {\gamma}^{(j)}_{(k_1,\ldots,k_t-1,\ldots k_m)}
\end{cases},
\end{align}
for all possible $m$-uples $(i_1,\ldots,i_m)$ and $(k_1,\ldots,k_m)$ and all $i_s$ and $k_t$ varying from $1$ to $m$, where
\begin{align}
\label{eq:system-f-g}
{\phi}^{(j)}_{(i_1, \ldots,i_m)}(\varepsilon) &= \mathrm{coeff}(f_j(\xvec; \varepsilon),x_1^{i_ 1}\cdots x_m^{i_m}), \\
{\gamma}^{(j)}_{(k_1,,\ldots k_m)} &= \mathrm{coeff}(g_j(\xvec),x_1^{i_ 1}\cdots x_m^{i_m}).
\end{align}
\end{theorem}

Theorem~\ref{theorem:existence-F-G} can be extended to the $(\dl, \dr, m, L, \dw)$ coupled ensemble in a straightforward manner.

We now give a necessary condition on the existence of $U(\xvec;\varepsilon)$ with diagonal $\Dvec$. We use the following definition.
{\definition
\label{definition:Sf-Sg}
For a vector function $\fvec(\xvec)=(f_1(\xvec), \ldots, f_m(\xvec))$, define the {\it coefficient sets} $\setf{1}, \ldots, \setf{m}$ as 
\begin{align}
\setf{j}&=\{(i_1,\ldots,i_m):\mathrm{coeff}(f_j(\xvec),x_1^{i_ 1}\cdots x_m^{i_m})\neq 0\},
\end{align}
for all $ j, \ 1\le j \le m$.
}

\begin{theorem}
\label{theorem:diagD}
Assume a diagonal matrix $\Dvec$. Then, the system of equations (\ref{eq:system}) exists if, for all $i$ from $1$ to $m$ 
\begin{align*}
&(i_1,\ldots,i_m)\in\set{f}{i}
\Leftrightarrow  (i_1,\ldots,i_i+1,\ldots,i_j-1,\ldots,i_m)\in\set{f}{j}\\
&(i_1,\ldots,i_m)\in\set{g}{i}\Leftrightarrow  (i_1,\ldots,i_i+1,\ldots,i_j-1,\ldots,i_m)\in\set{g}{j}
\end{align*}
for some values of $j$.
\end{theorem}
\begin{proposition} 
\label{prop:prop} For the $(\dl,\dr,m)$ nonbinary LDPC code ensemble, if $\Dvec$ is a diagonal matrix, the solution of (\ref{eq:system}) does not exist.
\end{proposition} 

The consequence of Proposition~\ref{prop:prop} is that the potential function as defined in \cite{pfister-itw} does not exist for nonbinary LDPC codes. However, we can prove the following proposition.
\begin{proposition}
A positive symmetric matrix $\Dvec$ is sufficient for the existence of a solution of (\ref{eq:system}).
\end{proposition}
\begin{table}[!t]
\addtolength{\tabcolsep}{-0.4mm}
\caption{DE thresholds for nonbinary SC-LDPC codes}
\vspace{-6ex}
\begin{center}\begin{tabular}{cccccccc}
\toprule
Ensemble& Rate  & $\epsilon^1_{\text{BP}}$ & $\epsilon^3_{\text{BP}}$ & $\epsilon^5_{\text{BP}}$ & $\epsilon^8_{\text{BP}}$ & $\epsilon_{\text{MAP}}$ & $\delta_{\rm{Sh}}$\\
\otoprule
$(3,6)$   & $1/2$  &0.4880& 0.4978  & 0.4995 & 0.4998 & 0.4999 & 0.0002\\[0.5mm]
$(3,9)$   & $2/3$  &0.3196 & 0.3307  & 0.3328 & 0.3331 & 0.3332 & 0.0002\\[0.5mm]
$(3,12)$  & $3/4$ &0.2372 & 0.2476 & 0.2495 & 0.2497 &0.2499 & 0.0003\\[0.5mm]
$(3,15)$  & $4/5$ &0.1886 & 0.1978 & 0.1995 & 0.1996 & 0.1999 & 0.0004\\[0.5mm]
\bottomrule
\end{tabular} \end{center}
\label{Tab:DEThreshold} 
\vspace{-5ex}
\end{table}

Thus, for nonbinary LDPC codes we can define the potential functions as in Definition~\ref{def:PotentialFunction} and~\ref{def:PotentialFunctionVec}, which are then used to prove threshold saturation.

\section{Numerical Results}

For the numerical results, we consider the $(\dl, \dr, m, L)$ coupled ensemble defined in~\cite{KuRiUr11}, properly extended to the nonbinary case. In Table~\ref{Tab:DEThreshold} we give the BP thresholds for several ensembles and $m=1,3,5$ and $8$, denoted by $\epsilon^1_{\text{BP}}$, $\epsilon^3_{\text{BP}}$, $\epsilon^5_{\text{BP}}$, and $\epsilon^8_{\text{BP}}$, respectively, for $L\rightarrow\infty$. It is observed that the threshold improves with $m$. In particular, a significant improvement is observed from $m=1$ (binary) to $m=3$. It is interesting to note that $\epsilon_{\text{BP}}$ approaches the Shannon limit as $m$ increases (the last column of the table gives the gap to the Shannon limit for the coupled ensembles with $m=8$, $\delta_{\rm{Sh}}$). We also observed for all values of $m$ that the BP threshold tends to the MAP threshold $\epsilon_{\text{MAP}}$ with  increasing values of $L$, suggesting threshold saturation to the MAP threshold. As an 
example, we report in the table the MAP threshold for $m=8$. 

In Fig.~\ref{fig:BER} we give bit error rate (BER) results for several nonbinary SC-LDPC codes with $m=3$, $L=65$ and codeword length $N=100$K bits. The code rate is $R=\left(1-\frac{\dl}{\dr}\right)-\frac{1}{L}$, where $\frac{1}{L}$ is the rate loss due to finite $L$. As a comparison, we also plot the performance for $m=1$ (binary code). In agreement with the DE results, the nonbinary SC-LDPC codes outperform their binary counterparts. 

\section{Conclusions}
\label{sec:Conclusions}

We proved threshold saturation for nonbinary SC-LDPC codes, when transmission takes place over the BEC, extending the proof in \cite{pfister-itw} to accommodate nonbinary SC-LDPC codes. We showed that nonbinary SC-LDPC codes achieve better BP threshold than their binary counterparts. Interestingly, the BP threshold approaches the Shannon limit with increasing values of $m$, suggesting that capacity can be achieved with nonbinary SC-LDPC codes. Finite length performance results confirm that nonbinary SC-LDPC codes may perform better than binary codes for a given (binary) block length.
 \begin{figure}
  \centering{}
  \includegraphics[width=0.85\columnwidth]{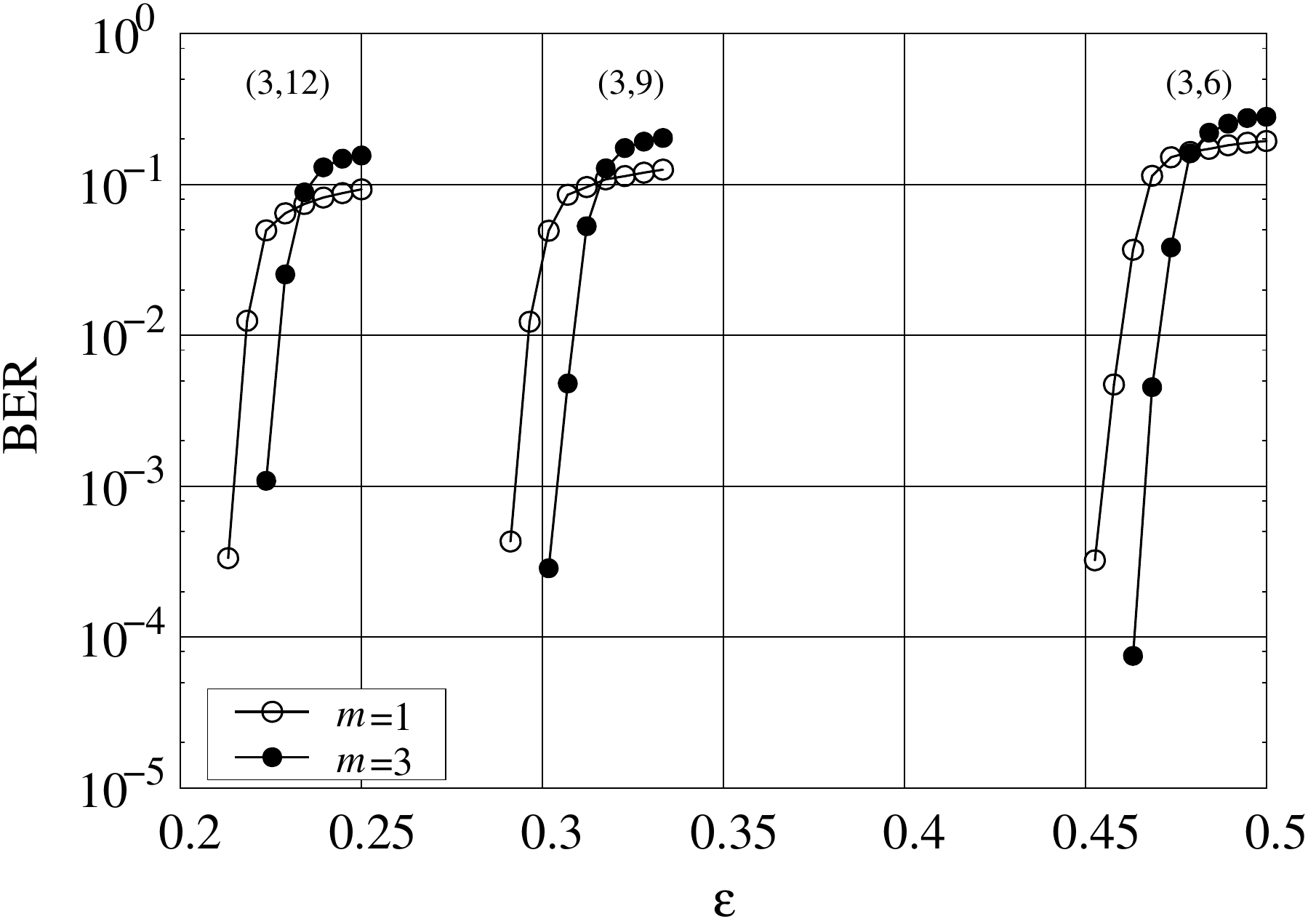} 
  \vspace{-0.3cm}
  \caption{BER performance of nonbinary SC-LDPC codes for $m=1$ and $3$.\label{f:BP36}}
  \vspace{-0.5cm}
 \label{fig:BER}
  \end{figure}


\bibliographystyle{IEEEtran}

\end{document}